\documentclass[12pt]{article}

\usepackage[T1]{fontenc}
\usepackage[utf8]{inputenc} 

\usepackage{upgreek} 

\usepackage[colorlinks=true,urlcolor=blue]{hyperref}

\usepackage{lmodern} 
\usepackage{multirow}
\usepackage{multicol}

\usepackage{graphicx} 
\usepackage{caption} 
\usepackage{subcaption} 
\usepackage{wrapfig} 
\usepackage{here} 

\usepackage{float}
\usepackage{array}

\usepackage{tikz} 
\usepackage{tikz-cd}
\usetikzlibrary{arrows,calc,chains,matrix,positioning,scopes} 

\usepackage[top=3.5cm, bottom=3.5cm, left=2.5cm, right=2.5cm]{geometry} 

\usepackage{scalerel}
\usepackage{stackengine}

\usepackage{yhmath}
\usepackage{amsthm} 
\usepackage{amsmath} 
\usepackage{amsfonts}
\usepackage{amssymb} 
\usepackage{mathrsfs} 
\usepackage{esint} 
\usepackage{empheq}
\usepackage{stmaryrd}
\usepackage{blkarray} 

\usepackage{cancel} 

\DeclareSymbolFont{UPM}{U}{eur}{m}{n}
\DeclareMathSymbol{\uppartial}{0}{UPM}{"40}

\newcommand{\hodge}{\mathop{}\mathopen{}\mathrm{{}^{*}}}

\newcommand{\id}{\mathop{}\mathopen{}\mathrm{Id}}

\newcommand{\opd}{\mathop{}\mathopen{}\mathrm{d}}

\newcommand{\abs}[1]{\left\lvert#1\right\rvert}

\newcommand{\ensemblenombre}[1]{\mathbb{#1}}

\newcommand{\R}{\ensemblenombre{R}}

\newcommand{\intervalle}[4]{\mathopen{#1}#2\mathclose{}\mathpunct{};#3\mathclose{#4}}

\newcommand{\intervalleentier}[2]{\intervalle{\llbracket}{#1}{#2}{\rrbracket}}

\newcommand{\trace}[1]{\mathrm{Tr}\mathopen{}\left(#1\right)}

\newcommand\reallywidetilde[1]{\ThisStyle{%
  \setbox0=\hbox{$\SavedStyle#1$}%
  \stackengine{-.1\LMpt}{$\SavedStyle#1$}{%
    \stretchto{\scaleto{\SavedStyle\mkern.2mu\sim}{.5\wd0}}{.4\ht0}%
  }{O}{c}{F}{T}{S}%
}}

\theoremstyle{definition}

\theoremstyle{plain}

\begin{document}

	\title{Geometric aspects of interpolating gauge fixing in Chern-Simons Theory}
	\date{}
	\author{Laurent \textsc{Gallot}, Philippe \textsc{Mathieu}, Éric \textsc{Pilon}, Frank \textsc{Thuillier}}
	\maketitle
\vspace{-1.2cm}
\begin{center}
{\it Laboratoire d'Annecy-le-Vieux de Physique Th\'eorique LAPTh,\\
 Universit\'e Savoie Mont Blanc, CNRS, BP 110, F-74941 Annecy, France}.
\end{center}

\vspace{2cm}

\begin{abstract}
In this article we investigate an interpolating gauge fixing procedure in $(4l+3)$-dimensional abelian Chern-Simons theory. We show that this interpolating gauge is related to the covariant gauge in a constant anisotropic metric. We compute the corresponding propagators involved in various expressions of the linking number in various gauges. We comment on the geometric interpretations of these expressions, clarifying how to pass from one interpretation to another.
\end{abstract}

\newpage

\section{Introduction}

The building blocks providing the observables of the $SU(2)$ Chern-Simons (CS) theory are traces of $SU(2)$ holonomies, a.k.a. Wilson loops. It is well-known \cite{W} that expectation values of these observables provide link invariants of the closed $3$-manifold $M$ on which the theory is considered. In the case of the $U\left(1\right)$ Chern-Simons theory these invariants all derive from linking numbers \cite{GT1}.

In the path integral formalism, the expectation value of a Wilson loop $W\left(\gamma\right)$ is written as:
\begin{equation}
\label{Observable}
\langle W\left(\gamma\right)\rangle
= \frac{1}{\mathscr{N}}\int_{\mathscr{H}}\mathscr{D}\!A\,
e^{2i\pi k \mathcal{S}_{{\mathrm{CS}}}\left[A\right]} \trace{\mathcal{P}e^{2i\pi\oint_{\gamma}A}} \, .
\end{equation}
where $\mathcal{S}_{\mathrm{\mathrm{CS}}}$ is the CS action, $\mathscr{H}$ the space of gauge fields and $\mathscr{N}$ a normalization. In the $SU(2)$ case we can pick in any gauge class a representative which is a global $su(2)$-valued $1$-form on $M$ so that the $SU(2)$ CS action can be written as:
\begin{equation}
\label{SU2action}
\mathcal{S}_{{\mathrm{CS}}}\left[A\right] = \frac{1}{8 \pi^2} \int_M Tr(A \wedge dA + \frac{2}{3} A \wedge A \wedge A ) \, .
\end{equation}
In the $U(1)$ case, because not every gauge class admits a global representative \cite{GT1}, the expression of the CS action is more complicated. In fact in that case the lagrangian turns out to be a square of so-called Deligne-Beilinson cohomology classes \cite{GT1}. Locally, that is to say in any contractible open subset $U$ of $M$, the $U(1)$ CS action can be written\footnote{Let us point out the difference of normalization between the abelian and non-abelian case: $\frac{1}{4 \pi^2}$ for the former and $\frac{1}{8 \pi^2}$ for the latter.} as:
\begin{equation}
\label{U1action}
\mathcal{S}_{{\mathrm{CS}}}\left[A\right] = \frac{1}{4 \pi^2} \int_U A \wedge dA \, ,
\end{equation}
and since any such contractible open subset is diffeomorphic to $\mathbb{R}^3$ it is not surprising to recover the usual expression of the $U(1)$ CS action on $\mathbb{R}^3$.

The CS action and the observables are topological in nature in the sense they are independent of any choice of metric, unlike the Yang-Mills or Maxwell actions. Moreover, as the CS action only depends on the class of the gauge field $A$, in the Quantum Field Theory context, which implies that $M = \mathbb{R}^3$, the computation of expectation values of observables as defined in equation \eqref{Observable} requires to fix the gauge so as to factor out the gauge redundancy. The gauge is usually chosen to be the covariant gauge which implies the introduction of the Euclidean metric in $\mathbb{R}^3$. At the end of the computation it can be checked that the expectation values do not depend on this metric.

\vspace{0.3cm}

From now on we focus on the $U(1)$ case returning the comments on the non-abelian case to the conclusion of the article.
The central quantity in the computation of expectation values (see \cite{GPT1}) is the two-point autocorrelator of the gauge field $A$. This quantity is gauge fixing dependant and so is the analytic expression of the linking number although its evaluation is not.

\vspace{0.3cm}

In this paper we explore a family of derivative linear gauges interpolating between commonly used gauges such as the covariant gauge previously explored in \cite{GPT1}, the Coulomb gauge and (in some sense explained below) the axial gauge. As this interpolating gauge requires the introduction of a metric, namely the Euclidean metric, the two-point autocorrelator of the gauge field depends on the metric although the expectation values of Wilson loops don't. Similar interpolating gauges, in Minkowski metrics, were considered in various contexts such as UV power counting and renormalisability properties in $4$D Yang-Mills theory \cite{BZ,DGLSSSTV} and vector supersymmetry in $3$D non-abelian CS theory \cite{LLSS}.

\vspace{0.3cm}

The aim of this paper is $2$-fold. 1. We wish to show how the interpolating gauge in the Euclidean metric can be related to the covariant gauge in a constant anisotropic metric in a natural way. 2. The various expressions of the linking number in various gauges have different geometric interpretations: in terms of Gauss' zenith/sphere in the covariant gauge, of transverse intersection in Coulomb gauge, of intertwining in the axial gauge. Using the interpolating gauge as well as point 1. we aim at clarifying how to pass from one interpretation to another.

\vspace{0.3cm}

We adopt a geometric formulation which makes the generalisation from $3$ to $4l+3$ dimensions more handy. Note that such a generalisation of the CS action is only possible in the $U(1)$ case. In particular, we deal with lagrangian {\em forms} instead of lagrangian {\em densities}, the correspondence between which being made explicit in section \ref{Conventions}. together with our notations and conventions. We then evaluate the two-point autocorrelator of the gauge field $A$ following two approaches in section \ref{Correlators}: one in the interpolating gauge using the Euclidean metric $\delta_{\mu\nu}$, the other using the covariant gauge in a constant anisotropic metric $\tilde{g}_{\mu\nu}$. In both cases the gauge fixed action is written as a quadratic form $\left(\mathcal{A},\mathcal{D}\mathcal{A}\right)$ with respect to $\delta_{\mu\nu}$, resp. $\reallywidetilde{\left(\mathcal{A},\widetilde{\mathfrak{D}}\mathcal{A}\right)}$ with respect to $\tilde{g}_{\mu\nu}$. The inversion of the differential operator $\mathcal{D}$, resp. $\mathfrak{D}$ provides the two-point autocorrelator. It is performed by means of the Fourier transform introduced in section \ref{Conventions}. The comparison between those two approaches leads to geometric interpretations. In section \ref{Limits} we explore various limits of the interpolating gauge recovering a number of familiar gauges. In section \ref{Observables} we illustrate our purpose by explicit expressions in the $3$ dimensional case which makes visualisation easier. There we supplement a discussion of the loop configurations which shall be ``forbidden'' to compute the link invariant of ambient isotopy which these loops belong to, in function of the gauge condition considered. We finish with a few remarks on topological gauge fixing. Detailed computations can be found in appendix.

\section{Conventions}
\label{Conventions}

\subsection{Fields}

We denote the vectors by lower case letters $n=n^{\mu}\partial_{\mu}$ and use capital or greek letters for the forms $\omega^{\left(r\right)} = \omega_{\mu_{1}\cdots\mu_{r}}\psi^{\mu_{1}}\wedge\cdots\wedge\psi^{\mu_{r}}$, with $\left(\psi^{\mu_{1}}\wedge\cdots\wedge\psi^{\mu_{r}}\right)_{\mu_{1},\cdots,\mu_{r}\in\intervalleentier{1}{4l+3}}$ a basis of the subspace of the $r$-forms of the exterior algebra $\bigwedge^{4l+3}(\mathbb{R}^{4l+3})$. Hence, a lowercase letter can be transformed into a capital one via the action of a metric on $\mathbb{R}^{4l+3}$. For the two metrics under consideration in this paper, the Euclidean one $\delta_{\mu\nu}$ and an anisotropic one $\widetilde{g}_{\mu\nu}$, one writes $N_{\mu}=\delta_{\mu\nu}n^{\nu}$ and $\widetilde{N}_{\mu}=\widetilde{g}_{\mu\nu}n^{\nu}$ for a given vector $n$. Note that although we adopt the notations of Differential Geometry the covariance involved in this article is just the one of $GL(\mathbb{R}^{4l+3})$.

We consider the so-called Levi-Civita symbol which is the totally antisymmetric tensor defined by:
\begin{equation}
\varepsilon_{\mu_{1}\cdots\mu_{4l+3}}=
\begin{cases}
+1 &\text{ if } \varepsilon\left(\left(\mu_{1}\cdots\mu_{4l+3}\right)\right)=+1 \\
0 &\text{ if } \exists\,\, i,j \in \intervalleentier{1}{\left(4l+3\right)}\quad | \quad\mu_{i} = \mu_{j} \\
-1 &\text{ if } \varepsilon\left(\left(\mu_{1}\cdots\mu_{4l+3}\right)\right)=-1
\end{cases}
\end{equation}
where $\varepsilon\left(\left(\mu_{1}\cdots\mu_{4l+3}\right)\right)$ is the signature of the permutation $\left(\mu_{1}\cdots\mu_{4l+3}\right)$.

Let $\omega^{\left(r\right)}=\frac{1}{r!}\omega_{\mu_{1}\cdots\mu_{r}}\psi^{\mu_{1}}\wedge\cdots\wedge\psi^{\mu_{r}}$ be a $r$-form. The Hodge $\widetilde{\hodge }$ of $\omega$ is defined with respect to $\widetilde{g}$ by:
\begin{equation}
\label{Hodge}
\widetilde{\hodge}\omega^{\left(r\right)}
=\frac{\sqrt{\abs{\widetilde{g}}}}{r!\left(\left(4l+3\right)-r\right)!}\widetilde{g}^{\mu_{1}\nu_{1}}\cdots \widetilde{g}^{\nu_{r}\mu_{r}}\varepsilon_{\nu_{1}\cdots\nu_{\left(4l+3\right)}}\omega_{\mu_{1}\cdots\mu_{r}}\psi^{\nu_{r+1}}\wedge\cdots\wedge\psi^{\nu_{\left(4l+3\right)}}
\end{equation}
with $\abs{\widetilde{g}} = \det\left(\widetilde{g}\right)$.

Note that since $4l+3$ is odd and the metrics which we consider are riemannian:
\begin{equation}
\widetilde{\hodge }\widetilde{\hodge }\omega^{\left(r\right)} = \omega^{\left(r\right)}
\end{equation}

Let $\eta^{\left(q\right)} = \frac{1}{q!}\eta_{\mu_{1}\cdots\mu_{q}}\psi^{\mu_{1}}\wedge\cdots\psi^{\mu_{q}}$, then the exterior product $\omega^{\left(r\right)}\wedge\eta^{\left(q\right)}$ is:
\begin{equation}
\omega^{\left(r\right)}\wedge\eta^{\left(q\right)}
= \frac{1}{r!q!}\omega_{\mu_{1}\cdots\mu_{r}}\eta_{\mu_{r+1}\cdots\mu_{r+q}}\psi^{\mu_{1}}\wedge\cdots\wedge\psi^{\mu_{r+q}}
\end{equation}

The correspondance between the lagrangian form $\mathcal{L}$ and the lagrangian density $L$ is then $\mathcal{L}=\hodge L$ so that $\int \mathcal{L} = \int L \sqrt{\abs{\tilde{g}}} d^{n}x$ according to equation \eqref{Hodge}.

\subsection{Fourier transform of forms w.r.t. a constant anisotropic metric}

In the following, the fields which we deal with are elements of the tensor product between smooth functions and the exterior algebra $\mathcal{C}^{\infty}\left(\R^{4l+3}_x\right)\otimes\bigwedge^*(\R^{4l+3})$. The Fourier transform is defined so as to send the set of fields in a copy of itself $\mathcal{C}^{\infty}\left(\R^{4l+3}_k\right)\otimes\bigwedge^*(\R^{4l+3})$. In particular this Fourier transform preserves form degrees.

To $\omega^{\left(r\right)} = \frac{1}{r!}\omega_{\mu_{1}\cdots\mu_{r}}\left(x\right)\psi^{\mu_{1}}_{x}\wedge\cdots\wedge\psi^{\mu_{r}}_{x}$ we associate the Fourier transform:
\begin{equation}
\widetilde{FT}\left[\omega^{\left(r\right)}\right]
= \frac{1}{r!}\widetilde{FT}\left[\omega_{\mu_{1}\cdots\mu_{r}}\right]\left(k\right)\psi^{\mu_{1}}_{k}\wedge\cdots\wedge\psi^{\mu_{r}}_{k}
\end{equation}
where:
\begin{equation}
\widetilde{FT}\left[\omega_{\mu_{1}\cdots\mu_{r}}\right]\left(k\right)
=\int_{\R^{4l+3}}\omega_{\mu_{1}\cdots\mu_{r}}\left(x\right)e^{-ik^{\mu}\tilde{g}_{\mu\nu}x^{\nu}}\opd^{4l+3}x
\end{equation}
We stress that the definition of a Fourier transform associated with $\tilde{g}$ relies heavily on the fact that this metric is constant albeit anisotropic .

To $\phi^{\left(r\right)} = \frac{1}{r!}\phi_{\mu_{1}\cdots\mu_{r}}\left(k\right)\psi^{\mu_{1}}_{k}\wedge\cdots\wedge\psi^{\mu_{r}}_{k}$ we associate the inverse Fourier transform:
\begin{equation}
\widetilde{FT}^{-1}\left[\phi^{\left(r\right)}\right]
= \frac{1}{r!}\widetilde{FT}^{-1}\left[\phi_{\mu_{1}\cdots\mu_{r}}\right]\left(x\right)\psi^{\mu_{1}}_{x}\wedge\cdots\wedge\psi^{\mu_{r}}_{x}
\end{equation}
where:
\begin{equation}
\widetilde{FT}^{-1}\left[\phi_{\mu_{1}\cdots\mu_{r}}\right]\left(x\right)
=\frac{1}{\left(2\pi\right)^{4l+3}}\int_{\R^{4l+3}}\phi_{\mu_{1}\cdots\mu_{r}}\left(k\right)e^{ik^{\mu}\widetilde{g}_{\mu\nu}x^{\nu}}\abs{\widetilde{g}}\opd^{4l+3}k
\end{equation}

With these definitions, we have:
\begin{equation}
\widetilde{FT}\left[\partial_{\mu}\right] = -i \widetilde{K}_{\mu}
\end{equation}

Convolution is defined as usual:
\begin{equation}
u\odot v\left(x\right) = \int_{\R^{4l+3}}u\left(y\right)v\left(x-y\right)\opd y
\end{equation}
so that $\widetilde{FT}$ turns convolutions into pointwise products:
\begin{equation}
\widetilde{FT}\left[u\odot v\right] = \widetilde{FT}\left[u\right]\cdot\widetilde{FT}\left[v\right]
\end{equation}

\section{Correlators}
\label{Correlators}

\subsection{Evaluation in Euclidean metric}

In this subsection, we consider the specific case of the Euclidean metric $\delta_{\mu\nu}$.

In $3$ dimensions link invariants between two loops $\gamma_{1}$ and $\gamma_{2}$ may be expressed in terms of the two-point autocorrelator of $A$ of the $U(1)$ CS Theory \cite{GPT1}. Gauge fixing is required in order to define this two-point autocorrelator, which depends on this gauge fixing, despite the fact that the link invariants themselves do not. In this article we explore the features of a family of linear derivative gauge fixings interpolating between the familiar covariant or Lorentz gauge function $\partial^{\mu} A_{\mu}$, the so-called Coulomb gauge function $\partial^{\mu} A_{\mu} -( n^{\mu} \partial_{\mu}) (n^{\nu} A_{\nu})$ (where $n$ is some constant vector normalised to $1$) and the gauge function $(n^{\mu} \partial_{\mu}) (n^{\nu} A_{\nu})$, which we name ``axial limit'' related to the so-called axial gauge function $n^{\mu} A_{\mu}$ in a sense explained further. The interpolating gauge fixing is parametrised by the gauge fixing function:
\begin{equation}
\mathcal{F}\left(A\right)=\alpha \left(\delta^{\mu\nu}\partial_{\mu}A_{\nu}\right) + \beta \left(n^{\nu}\partial_{\nu}\right)\left(n^{\mu}A_{\mu}\right)
\end{equation}
where $\alpha$ and $\beta$ are real numbers. This constraint is implemented by adding the Lagrange multiplier term $B\,\mathcal{F}\left(A\right)$ where $B$ is an auxiliairy field, to the CS lagrangian. We set for further convenience $\alpha + \beta = \gamma$. We have written here the usual $\partial^{\mu}A_{\mu}$ as $g^{\mu\nu}\partial_{\mu}A_{\nu}$ to underline the fact that $\partial_{\mu}$ is a natural covariant operator and thus requires a metric to act on $A_{\mu}$ which is also, as a $1$-form, naturally covariant. We restrict ourselves to the parameter range $\alpha\gamma \geq 0$. The generic subcase $\alpha \gamma > 0$ will appear as some smooth deformation of the isotropic case $\beta = 0$ which corresponds to the covariant gauge. The limit $\gamma = 0$ corresponds to the Coulomb gauge. The so-called axial limit $\alpha \rightarrow 0$, leads to recover the results in the axial gauge, although it is \textit{not} the axial gauge \textit{stricto sensu}.
The restriction $\alpha\gamma \geq 0$ on the parameter range will be commented in the conclusion.

In order to easily generalise to  $4l+3$ dimensions, it is convenient to translate the Lagrange constraint in geometric terms using exterior derivative, wedge product and Hodge dualisation:
\begin{equation}
\label{GF4l+3}
\mathcal{L}_{GF} = B  \wedge [ \alpha \, d \hodge  A + \beta N  \wedge \hodge  d \wedge N \hodge  A]
\end{equation}
where $\alpha$ and $\beta$ are submitted to the same constraint as above. Expression (\ref{GF4l+3}) shall be understood as follows: the operations of Hodge $\hodge$, wedge product and exterior derivative successively act from right to left on the whole forms on their respective right sides. For example,
\begin{equation}
B \wedge N  \wedge \hodge  d \wedge N \hodge  A
\Longleftrightarrow
B \wedge \langle N  \wedge \lbrace \hodge  \left[ d \left( \hodge  (N \wedge \hodge  A) \right) \right] \rbrace\rangle
\end{equation}
In $4l+3$ dimensions $A=A^{\left(2l+1\right)}$ is a $\left(2l+1\right)$-form, thus $B=B^{\left(2l\right)}$ is a $2l$-form and the above gauge-fixing constraint is incomplete as a large residual gauge invariance with respect to $B^{\left(2l\right)}$ is left. The latter can be fixed in its turn using a similar procedure with a $\left(2l-1\right)$-form of Lagrange multipliers $B^{\left(2l-1\right)}$ etc. in a cascading way. The complete gauge fixing lagrangian then read:
\begin{align}
\nonumber
\mathcal{L}_{GF}^{\left(4l+3\right)} =
&\alpha B^{\left(2l\right)} \wedge \opd \hodge  A^{\left(2l+1\right)} + \beta B^{\left(2l\right)} \wedge N\wedge\hodge \opd \hodge N\wedge \hodge A^{\left(2l+1\right)} \\
\nonumber
&+\alpha B^{\left(2l-1\right)} \wedge \opd \hodge  B^{\left(2l\right)} + \beta B^{\left(2l-1\right)} \wedge N\wedge\hodge \opd \hodge N\wedge\hodge B^{\left(2l\right)} \\
\nonumber
&+ \cdots \\
&+\alpha B^{\left(0\right)} \wedge \opd \hodge  B^{\left(1\right)} + \beta B^{\left(0\right)} \wedge N\wedge\hodge \opd \hodge N\wedge\hodge B^{\left(1\right)}
\end{align}
We now focus on the computation of this two-point autocorrelator of $A$ in the interpolating gauge considered. For this purpose we define $\mathcal{A} = \left(A^{\left(2l+1\right)},B^{\left(2l\right)},B^{\left(2l-1\right)},\cdots,B^{\left(0\right)}\right)$. The complete action which we deal with can be written as a scalar product:
\begin{equation}
\frac{1}{2}\left(\mathcal{A},\mathfrak{D}\mathcal{A}\right):= \int_{\R^{4l+3}}\mathcal{L}_{\mathrm{CS}}+\mathcal{L}_{GF}.
\end{equation}
In this equation, $\mathfrak{D}$ is the $\left(2l+2\right) \times \left(2l+2\right)$ matrix differential operator:
\begin{equation}
\mathfrak{D}=
\begin{psmallmatrix}
\left(\hodge \opd\right)^{\left(2l+1,2l+1\right)} & -\Psi^{\left(2l+1,2l\right)} & 0 & 0 &  &  &  &  &  &  \\
\Xi^{\left(2l,2l+1\right)} & 0 & \Psi^{\left(2l,2l-1\right)} & 0 &  &  &  &  &  &  \\
0 & \Xi^{\left(2l-1,2l\right)} & 0 & -\Psi^{\left(2l-1,2l-2\right)} &  &  &  &  &  &  \\
0 & 0 & \Xi^{\left(2l-2,2l-1\right)} & 0 &  &  &  &  &  &  \\
 &  &  &  &  &  &  &  &  &  \\
 &  &  &  &  & 0 & -\Psi^{\left(j+1,j\right)} & 0 &  &  \\
 &  &  &  &  & \Xi^{\left(j,j+1\right)} & 0 & \Psi^{\left(j,j-1\right)} &  &  \\
 &  &  &  &  & 0 & \Xi^{\left(j-1,j\right)} & 0 &  &  \\
 &  &  &  &  &  &  &  &  &  \\
 &  &  &  &  &  &  &  & 0 & -\Psi^{\left(1,0\right)}  \\
 &  &  &  &  &  &  &  & \Xi^{\left(0,1\right)} & 0
\end{psmallmatrix}
\end{equation}
where:
\begin{equation}
\Xi=\left(\alpha \hodge  \opd \hodge  +\beta \hodge N\wedge\hodge \opd \hodge N\wedge\hodge \right)
\end{equation}
and:
\begin{equation}
\Psi=\left(\alpha \opd +\beta N\wedge\hodge \opd \hodge N\wedge \right).
\end{equation}
The right superscript indicates the degree of the form on which the operator acts, the left superscript indicates the degree of the resulting form. Note that $\Xi \neq \pm \hodge \Psi\hodge $ since the degree on which each operates is not the same.

With respect to the conventions introduced in section \ref{Conventions} the Fourier transform of $\mathfrak{D}$ is the $\left(2l+2\right) \times \left(2l+2\right)$ matrix multiplicative operator $\hat{\mathfrak{D}}$:
\begin{equation}
\hat{\mathfrak{D}}=
\begin{psmallmatrix}
\left(\hodge \hat{\opd}\right)^{\left(2l+1,2l+1\right)} & -\hat{\Psi}^{\left(2l+1,2l\right)} & 0 & 0 &  &  &  &  &  &  \\
\hat{\Xi}^{\left(2l,2l+1\right)} & 0 & \hat{\Psi}^{\left(2l,2l-1\right)} & 0 &  &  &  &  &  &  \\
0 & \hat{\Xi}^{\left(2l-1,2l\right)} & 0 & -\hat{\Psi}^{\left(2l-1,2l-2\right)} &  &  &  &  &  &  \\
0 & 0 & \hat{\Xi}^{\left(2l-2,2l-1\right)} & 0 &  &  &  &  &  &  \\
 &  &  &  &  &  &  &  &  &  \\
 &  &  &  &  & 0 & -\hat{\Psi}^{\left(j+1,j\right)} & 0 &  &  \\
 &  &  &  &  & \hat{\Xi}^{\left(j,j+1\right)} & 0 & \hat{\Psi}^{\left(j,j-1\right)} &  &  \\
 &  &  &  &  & 0 & \hat{\Xi}^{\left(j-1,j\right)} & 0 &  &  \\
 &  &  &  &  &  &  &  &  &  \\
 &  &  &  &  &  &  &  & 0 & -\hat{\Psi}^{\left(1,0\right)}  \\
 &  &  &  &  &  &  &  & \hat{\Xi}^{\left(0,1\right)} & 0
\end{psmallmatrix}
\end{equation}
where:
\begin{align}
&\hat{\opd}^{\left(r+1,r\right)}
=i K\wedge\\
&\hat{\Xi}^{\left(r,r+1\right)}
=i\hodge Q\wedge\hodge \\
&\hat{\Psi}^{\left(r+1,r\right)}
=i Q\wedge
\end{align}
with:
\begin{equation}
Q_{\mu} = \delta_{\mu\nu}q^{\nu}
\end{equation}
and:
\begin{equation}
q = \alpha k + \beta \left(n\cdot k\right)n = \alpha k_{\perp}+\gamma k_{\parallel}
\end{equation}
and $\gamma = \alpha+\beta$.

We are looking for a right-inverse operator from the point of view of the convolution in the direct space:
\begin{equation}
\mathfrak{D} \odot \mathfrak{P} \left(x\right)=\delta\left(x\right)
\end{equation}
In Fourier space this reads:
\begin{equation}
\hat{\mathfrak{D}}\hat{\mathfrak{P}}=\id
\end{equation}
with:
\begin{equation}
\hat{\mathfrak{P}}=\left(\hat{\mathfrak{P}}^{\left(2l+2-i,2l+2-j\right)}_{i,j}\right)_{1 \leq i,j\leq 2l+2}
\end{equation}
The procedure of inversion is detailed in the appendix.

Let us focus on the component $\hat{\mathfrak{P}}^{\left(2l+1,2l+1\right)}_{1,1}$ since it is the only one that matters in the computation of the link invariant:
\begin{equation}
\hat{\mathfrak{P}}^{\left(2l+1,2l+1\right)}_{1,1}
=-i\hodge \frac{Q}{k \cdot q}\wedge
\end{equation}
Be careful that this equation describes actually an action on $2l+1$-forms: the Hodge star does not act on $Q$ but on $Q\wedge\omega^{\left(2l+1\right)}$.

We can rewrite $\hat{\mathfrak{P}}^{\left(2l+1,2l+1\right)}_{1,1}$ in terms of components:
\begin{equation}
\left(\hat{\mathfrak{P}}^{\left(2l+1,2l+1\right)}_{1,1}\right)_{\mu_{1}\cdots\mu_{2l+1}\nu_{1}\cdots\nu_{2l+1}}
=-\frac{i}{\left(2l+1\right)!}
\varepsilon_{\mu_{1}\cdots\mu_{2l+1}\nu_{1}\cdots\nu_{2l+1}\rho}
\frac{\alpha k_{\perp}^{\rho}+\gamma k_{\parallel}^{\rho}}{\alpha k_{\perp}^{2}+\gamma k_{\parallel}^{2}}
\end{equation}
where we took the convention $k_{\parallel} =\left(k^{1},0,\cdots,0\right)$ and $k_{\perp}=\left(0,k^{2},\cdots,k^{4l+3}\right)$. Note that the normalisation $\frac{1}{\left(2l+1\right)!}$ comes from the $\hodge $ applied to a $2l+2$-form.

By inverse Fourier transform (see details in appendix), we finally find:
\begin{align}
\mathfrak{P}_{\mu_{1}\cdots\mu_{2l+1}\nu_{1}\cdots\nu_{2l+1}}\left(x\right)
=\frac{1}{\left(2l+1\right)!}
\frac{\Gamma\left(\frac{4l+3}{2}\right)}{2\pi^{\frac{4l+3}{2}}}
\frac{1}{\left(\gamma\alpha^{4l+2}\right)^{\frac{1}{2}}}
\varepsilon_{\mu_{1}\cdots\mu_{2l+1}\nu_{1}\cdots\nu_{2l+1}\rho}
\frac{x^{\rho}}{\left(\frac{1}{\gamma}x_{\parallel}^{2}+\frac{1}{\alpha}x_{\perp}^{2}\right)^{\frac{4l+3}{2}}}
\end{align}

We can rewrite this formula in terms of the distance $\widetilde{\abs{\,\cdot\,}}$ defined by the anisotropic metric:
\begin{equation}
\widetilde{g}_{\mu\nu}=
\begin{pmatrix}
\frac{1}{\gamma} & 0 & \cdots & \cdots & 0 \\
0 & \frac{1}{\alpha} & 0 & \cdots & 0 \\
\vdots & & \ddots & & \vdots \\
\vdots & & & \ddots & 0 \\
0 & \cdots & \cdots & 0 & \frac{1}{\alpha} \\
\end{pmatrix}
\end{equation}
which gives:
\begin{align}
\mathfrak{P}_{\mu_{1}\cdots\mu_{2l+1}\nu_{1}\cdots\nu_{2l+1}}\left(x\right)
=\frac{1}{\left(2l+1\right)!}
\frac{\Gamma\left(\frac{4l+3}{2}\right)}{2\pi^{\frac{4l+3}{2}}}
\frac{1}{\left(\gamma\alpha^{4l+2}\right)^{\frac{1}{2}}}
\varepsilon_{\mu_{1}\cdots\mu_{2l+1}\nu_{1}\cdots\nu_{2l+1}\rho}
\frac{x^{\rho}}{\quad\widetilde{\abs{x}}^{4l+3}}
\end{align}

This two-point autocorrelator is reminiscent from the well-known Gauss linking formula according to the anisotropic metric $\widetilde{g}$. This will be enlightened in the next subsection.

\subsection{Evaluation in anisotropic metric}

The interpolating gauge function may be written using the inverse metric
$\widetilde{g}^{\mu\nu}$. In the 3-dimensional case it reads:
\begin{equation}
\mathcal{F}\left(A\right)=\alpha \left(\delta^{\mu\nu}\partial_{\mu}A_{\nu}\right) +
\beta \left(n^{\nu}\partial_{\nu}\right)\left(n^{\mu}A_{\mu}\right)
= \widetilde{g}^{\mu\nu}\partial_{\mu}A_{\nu},
\end{equation}
$n$ being a fixed direction.
More generally in $4l+3$ dimensions the gauge fixing lagrangian reads:
\begin{align}
\nonumber
\widetilde{\mathcal{L}}_{GF} =
& B^{\left(2l\right)} \wedge \opd \widetilde{\hodge } A^{\left(2l+1\right)} \\
\nonumber
&+ B^{\left(2l-1\right)} \wedge \opd \widetilde{\hodge } B^{\left(2l\right)} \\
\nonumber
&+ \cdots \\
&+ B^{\left(0\right)} \wedge \opd \widetilde{\hodge } B^{\left(1\right)}
\end{align}
where the $B^{\left(k\right)}$s are auxiliary fields. It corresponds to
the ``covariant'' gauge yet in the anisotropic metric $\widetilde{g}$.
We now compute the two-point autocorrelator of $A$ in this gauge.
With $\mathcal{A} =
\left(A^{\left(2l+1\right)},B^{\left(2l\right)},B^{\left(2l-1\right)},\cdots,B^{\left(0\right)}\right)$, the
gauge fixed action reads:
\begin{equation}
\frac{1}{2}\reallywidetilde{\left(\mathcal{A}, \mathcal{\widetilde{D}A}\right)}
:= \frac{1}{2} \int_{\R^{4l+3}} \mathcal{L}_{\mathrm{CS}}+\widetilde{\mathcal{L}}_{GF}
\end{equation}
with $\widetilde{D}$ represented by the $\left(2l+2\right) \times \left(2l+2\right)$ matrix:
\begin{equation}
\begin{psmallmatrix}
\left(\widetilde{\hodge }\opd\right)^{\left(2l+1,2l+1\right)} & -\opd^{\left(2l+1,2l\right)} & 0 & 0 &  &  &  &  &  &  \\
\widetilde{\Phi}^{\left(2l,2l+1\right)} & 0 & \opd^{\left(2l,2l-1\right)} & 0 &  &  &  &  &  &  \\
0 & \widetilde{\Phi}^{\left(2l-1,2l\right)} & 0 & -\opd^{\left(2l-1,2l-2\right)} &  &  &  &  &  &  \\
0 & 0 & \widetilde{\Phi}^{\left(2l-2,2l-1\right)} & 0 &  &  &  &  &  &  \\
 &  &  &  &  &  &  &  &  &  \\
 &  &  &  &  & 0 & -\opd^{\left(j+1,j\right)} & 0 &  &  \\
 &  &  &  &  & \widetilde{\Phi}^{\left(j,j+1\right)} & 0 & \opd^{\left(j,j-1\right)} &  &  \\
 &  &  &  &  & 0 & \widetilde{\Phi}^{\left(j-1,j\right)} & 0 &  &  \\
 &  &  &  &  &  &  &  &  &  \\
 &  &  &  &  &  &  &  & 0 & -\opd^{\left(1,0\right)}  \\
 &  &  &  &  &  &  &  & \widetilde{\Phi}^{\left(0,1\right)} & 0
\end{psmallmatrix}
\end{equation}
with:
\begin{equation}
\widetilde{\Phi}=\widetilde{\hodge } \opd \widetilde{\hodge }
\end{equation}
The conventions on the superscripts are the same as the ones in the previous subsection. Note that $\Phi \neq \pm \widetilde{\hodge }\opd\widetilde{\hodge }$ since the degree on which each operates is not the same.

If we perform a Fourier transformation with respect of the metric $\widetilde{g}$ of $\widetilde{D}$ then we obtain an operator $\hat{\widetilde{D}}$ represented by the $\left(2l+2\right)\times\left(2l+2\right)$ matrix:
\begin{align}
\begin{psmallmatrix}
\left(\widetilde{\hodge }\hat{\opd}\right)^{\left(2l+1,2l+1\right)} & -\hat{\opd}^{\left(2l+1,2l\right)} & 0 & 0 &  &  &  &  &  &  \\
\hat{\widetilde{\Phi}}^{\left(2l,2l+1\right)} & 0 & \hat{\opd}^{\left(2l,2l-1\right)} & 0 &  &  &  &  &  &  \\
0 & \hat{\widetilde{\Phi}}^{\left(2l-1,2l\right)} & 0 & -\hat{\opd}^{\left(2l-1,2l-2\right)} &  &  &  &  &  &  \\
0 & 0 & \hat{\widetilde{\Phi}}^{\left(2l-2,2l-1\right)} & 0 &  &  &  &  &  &  \\
 &  &  &  &  &  &  &  &  &  \\
 &  &  &  &  & 0 & -\hat{\opd}^{\left(j+1,j\right)} & 0 &  &  \\
 &  &  &  &  & \hat{\widetilde{\Phi}}^{\left(j,j+1\right)} & 0 & \hat{\opd}^{\left(j,j-1\right)} &  &  \\
 &  &  &  &  & 0 & \hat{\widetilde{\Phi}}^{\left(j-1,j\right)} & 0 &  &  \\
 &  &  &  &  &  &  &  &  &  \\
 &  &  &  &  &  &  &  & 0 & -\hat{\opd}^{\left(1,0\right)}  \\
 &  &  &  &  &  &  &  & \hat{\widetilde{\Phi}}^{\left(0,1\right)} & 0
\end{psmallmatrix}
\end{align}
with:
\begin{align}
&\hat{\opd}^{\left(r+1,r\right)}
=-i \widetilde{K}\wedge\\
&\hat{\widetilde{X}}^{\left(r,r+1\right)}
=-i\widetilde{\hodge }\widetilde{K}\wedge\widetilde{\hodge }\\
\end{align}

We are looking for a right-inverse operator from the point of view of the convolution in the direct space:
\begin{equation}
\widetilde{D} \odot \widetilde{P} \left(x\right)=\delta\left(x\right)
\end{equation}
thus satisfying in Fourier space:
\begin{equation}
\hat{\widetilde{D}}\hat{\widetilde{P}}=\id
\end{equation}
with:
\begin{equation}
\hat{\widetilde{P}}=\left(\hat{\widetilde{P}}^{\left(2l+2-i,2l+2-j\right)}_{i,j}\right)_{1 \leq i,j\leq 2l+2}
\end{equation}
which leads to the system given and solved in appendix.

Finally:
\begin{equation}
\hat{\widetilde{P}}^{\left(2l+1,2l+1\right)}_{1,1}
=-i\widetilde{\hodge }\frac{\widetilde{K}}{k\widetilde{\cdot}k}\wedge
\end{equation}
which is perfectly analogous to our previous result:
\begin{equation}
\hat{\mathfrak{P}}^{\left(2l+1,2l+1\right)}_{1,1}
=-i\hodge \frac{Q}{k\cdot q}\wedge
\end{equation}
up to a dissymmetry between $q$ and $k$ and with $\widetilde{\cdot}$ and $\widetilde{\hodge }$ instead of $\cdot$ and $\hodge $.

We can rewrite $\hat{\widetilde{P}}^{\left(2l+1,2l+1\right)}_{1,1}$ in terms of components:
\begin{equation}
\left(\hat{\widetilde{P}}^{\left(2l+1,2l+1\right)}_{1,1}\right)_{\mu_{1}\cdots\mu_{2l+1}\nu_{1}\cdots\nu_{2l+1}}
=-\frac{i\sqrt{\abs{\widetilde{g}}}}{\left(2l+1\right)!}
\varepsilon_{\mu_{1}\cdots\mu_{2l+1}\nu_{1}\cdots\nu_{2l+1}\rho}
\frac{k^{\rho}}{\frac{1}{\alpha} k_{\perp}^{2}+\frac{1}{\gamma} k_{\parallel}^{2}}
\end{equation}
and by inverse Fourier transform, one finds finally:
\begin{equation}
\widetilde{P}_{\mu_{1}\cdots\mu_{2l+1}\nu_{1}\cdots\nu_{2l+1}}\left(x\right)
=\frac{1}{\left(2l+1\right)!}\frac{\Gamma\left(\frac{4l+3}{2}\right)}{2\pi^{\frac{4l+3}{2}}}
\frac{1}{\gamma\alpha^{4l+2}}
\varepsilon_{\mu_{1}\cdots\mu_{2l+1}\nu_{1}\cdots\nu_{2l+1}\rho}
\frac{x^{\rho}}{\quad\widetilde{\abs{x}}^{4l+3}}
\end{equation}
which is not exactly the same result as the $\mathfrak{P}$ found in the previous part because of the normalisation factor coming from the Hodge star operation in $\hat{\widetilde{P}}$. However, the normalisation factor is absorbed in the computations of observables:
\begin{equation}
\left(\mathcal{J}_{1},\mathfrak{P}\mathcal{J}_{2}\right)=\reallywidetilde{\left(\mathcal{J}_{1},\widetilde{P}{\mathcal{J}_{2}} \right)}
\end{equation}
by definition, so this operation contains a Hodge star and $\hat{\widetilde{P}}$ too and we saw in section \ref{Conventions} that double Hodge star is the identity. Both approaches are thus equivalent at the level of observables.

Note that a more general interpolating gauge can be generated by taking the covariant gauge with respect to a totally generic constant anisotropic metric. This more general case would be solved by choosing one of the two methods previously proposed: either in the interpolating gauge in the Euclidean metric, or in the covariant gauge in the generic constant anisotropic metric. This can be achieved mainly because our definition of the Fourier transform has a meaning for any constant anisotropic metric.

As a final note let us point out that we chose to consider metric changes: switching from the Euclidean metric $\delta_{\mu \nu}$ to the constant anisotropic metric $\widetilde{g}_{\mu\nu}$. Instead we could have considered the active $GL(\mathbb{R}^{4l+3})$ transformation of $\mathbb{R}^{4l+3}$ which transforms (via pullback or pushforward) the Euclidean metric into the anisotropic one. However it appeared to us that this approach would be more cumbersome since then we would also have to take gauge field transformations into account.

\section{Limits}
\label{Limits}

Well-known formulas such as Gauss density are recovered from specific limits of
this two-point autocorrelator.
The two-point autocorrelator in Euclidean metric is the most convenient one to study the limits:
\begin{equation}\label{start}
\mathfrak{P}_{\mu_{1}\cdots\mu_{2l+1}\nu_{1}\cdots\nu_{2l+1}}\left(x\right)
=\frac{1}{\left(2l+1\right)!}
\frac{\Gamma\left(\frac{4l+3}{2}\right)}{2\pi^{\frac{4l+3}{2}}}
\frac{1}{\left(\gamma\alpha^{4l+2}\right)^{\frac{1}{2}}}
\varepsilon_{\mu_{1}\cdots\mu_{2l+1}\nu_{1}\cdots\nu_{2l+1}\rho}
\frac{x^{\rho}}{\quad\widetilde{\abs{x}}^{4l+3}}
\end{equation}

\subsection{Covariant limit}

The propagator in the covariant gauge
is recovered in the limit
$\beta \rightarrow 0$ and $\alpha \rightarrow 1$, thus $\gamma \rightarrow 1$,
whereby we readily find:
\begin{equation}\label{genalphgam}
\mathfrak{P}_{\mu_{1}\cdots\mu_{2l+1}\nu_{1}\cdots\nu_{2l+1}}\left(x\right)
\underset{\substack{\alpha\to 1\\ \gamma\to 1}}{\longrightarrow}
\frac{1}{\left(2l+1\right)!}
\frac{\Gamma\left(\frac{4l+3}{2}\right)}{2\pi^{\frac{4l+3}{2}}}
\varepsilon_{\mu_{1}\cdots\mu_{2l+1}\nu_{1}\cdots\nu_{2l+1}\rho}
\frac{x^{\rho}}{\abs{x}^{4l+3}}
\end{equation}
The latter also coincides with the well-known Gauss density \cite{GPT1}.

\vspace{0.3cm}

\noindent
The Coulomb gauge and axial limit are recovered as distributional limits a little more subtle to take.

\subsection{Coulomb limit}

The propagator in the Coulomb gauge $\alpha=1,\gamma=0$.
is indeed recovered from eq. (\ref{genalphgam}) in the limit $\gamma \rightarrow 0$. To see this,
let us consider $\mathfrak{P}_{\mu_{1}\cdots\mu_{2l+1}\nu_{1}\cdots\nu_{2l+1}}$
acting on a test function $\phi$:
\begin{align}
\langle \mathfrak{P}_{\mu_{1}\cdots\mu_{2l+1}\nu_{1}\cdots\nu_{2l+1}},\phi \rangle
&=\frac{1}{\left(2l+1\right)!}
\frac{\Gamma\left(\frac{4l+3}{2}\right)}{2\pi^{\frac{4l+3}{2}}}\nonumber\\
&\varepsilon_{\mu_{1}\cdots\mu_{2l+1}\nu_{1}\cdots\nu_{2l+1}\rho}
\int_{\R^{4l+2}} \frac{d^{4l+2} x_{\perp}}{\sqrt{\alpha}^{4l+2}}
\int_{\R} \frac{d x_{\parallel}}{\sqrt{\gamma}}
\frac{x^{\rho}}{\left(\frac{1}{\gamma}x_{\parallel}^{2}+\frac{1}{\alpha}x_{\perp}^{2}\right)^{\frac{4l+3}{2}}}
\phi\left(x_{\parallel},x_{\perp}\right)
\end{align}
Let us rescale $x_{\parallel}=\sqrt{\gamma} \, y$.
The limit $\gamma \to 0$ of the integral over $y$ is controled by
Lebesgue's theorem of dominated convergence:
\begin{align}
\int_{\R} dy
\frac{\sqrt{\gamma} y \, n^{\rho} + x^{\rho}_{\perp}}
{\left( y^{2} + \frac{1}{\alpha} x_{\perp}^{2} \right)^{\frac{4l+3}{2}}}
\phi \left( \sqrt{\gamma} y , x_{\perp} \right)
&\underset{\substack{\gamma\to 0}}{\longrightarrow}
\sqrt{\pi} \frac{(2l)!}{\Gamma\left( \frac{4l+3}{2} \right)} \,
\left( \frac{\sqrt{\alpha}}{|x_{\perp}|} \right)^{4l+2} \,
x^{\rho}_{\perp} \, \phi \left( 0, x_{\perp} \right)
\notag
\end{align}
We obtain:
\begin{equation}
\label{coulombprop}
\mathfrak{P}_{\mu_{1}\cdots\mu_{2l+1}\nu_{1}\cdots\nu_{2l+1}}
\underset{\substack{\alpha\to 0}}{\longrightarrow}
\frac{1}{\left(2l+1\right)}
\frac{1}{2\pi^{2l+1}}
\varepsilon_{\mu_{1}\cdots\mu_{2l+1}\nu_{1}\cdots\nu_{2l+1}\rho}
\frac{x^{\rho}_{\perp}}{|x_{\perp}|^{4l+2}}
\delta^{\left(1\right)}\left(x_{\parallel}\right)
\end{equation}
which indeed coincides with the propagator directly computed in the Coulomb gauge
$\alpha=1,\gamma=0$.

\subsection{Axial limit}

The propagator computed directly for the gauge-fixing $\alpha=0,\gamma=1$ is recovered from  eq. (\ref{genalphgam}) in the limit $\alpha \rightarrow 0$.
To see this let us consider $\mathfrak{P}_{\mu_{1}\cdots\mu_{2l+1}\nu_{1}\cdots\nu_{2l+1}}$ acting
on a test function $\phi$:
\begin{align}
\langle \mathfrak{P}_{\mu_{1}\cdots\mu_{2l+1}\nu_{1}\cdots\nu_{2l+1}},\phi \rangle
&=\frac{1}{\left(2l+1\right)!}
\frac{\Gamma\left(\frac{4l+3}{2}\right)}{2\pi^{\frac{4l+3}{2}}}\nonumber\\
&\varepsilon_{\mu_{1}\cdots\mu_{2l+1}\nu_{1}\cdots\nu_{2l+1}\rho}
\int_{\R} \frac{d x_{\parallel}}{\sqrt{\gamma}}
\int_{\R^{4l+2}} \frac{d^{4l+2} x_{\perp}}{\sqrt{\alpha}^{4l+2}}
\frac{x^{\rho}}{\left(\frac{1}{\gamma}x_{\parallel}^{2}+\frac{1}{\alpha}x_{\perp}^{2}\right)^{\frac{4l+3}{2}}}
\phi\left(x_{\parallel},x_{\perp}\right)
\end{align}
Let us rescale $x^{\rho}_{\perp}=\sqrt{\alpha} \, y_{\perp}^{\rho}$.
The limit $\alpha \to 0$ of the integral over $y$ is again controled by Lebesgue's theorem of dominated
convergence:
\begin{align}
\int_{\R^{4l+2}} d^{4l+2} y_{\perp} \,
\frac{x_{\parallel} \, n^{\rho} + \sqrt{\alpha} \, y_{\perp}^{\rho}}
{\left( \frac{1}{\gamma}x_{\parallel}^{2}+y_{\perp}^{2}\right)^{\frac{4l+3}{2}}}
\phi\left(x_{\parallel},x_{\perp}\right)
&\underset{\substack{\alpha\to 0}}{\longrightarrow}
\frac{\pi^{\frac{4l+3}{2}}}{\Gamma \left( \frac{4l+3}{2} \right)} \,
\frac{x_{\parallel}}{|x_{\parallel}|} \, n^{\rho} \, \phi \left( x_{\parallel},0 \right)
\end{align}
We get:
\begin{equation}
\label{axialprop}
\mathfrak{P}_{\mu_{1}\cdots\mu_{2l+1}\nu_{1}\cdots\nu_{2l+1}}
=\frac{1}{2 \, \left(2l+1\right)!}
\varepsilon_{\mu_{1}\cdots\mu_{2l+1}\nu_{1}\cdots\nu_{2l+1}\rho}
\frac{x_{\parallel}}{\abs{x_{\parallel}}} \, n^{\rho}
\delta^{\left(4l+2\right)}\left(x_{\perp}\right)
\end{equation}
This two-point autocorrelator indeed coincides with the correlator in the gauge
$\alpha=0,\gamma=1$. It also happens to coincide with the expression of the two-point autocorrelator of $A$ in the so-called axial gauge.
A few comments are in order in this respect. The above observation may seem striking
at first since the axial gaauge is {\em not} a derivative gauge.
Yet it is not a complete
surprise as \textit{e.g.} in $3$ dimensions a gauge field $A$ fulfilling the axial gauge
condition $n^{\mu} A_{\mu} = 0$ fulfils \textit{a fortiori}
$(n^{\nu} \partial_{\nu})\, ( n^{\mu} A_{\mu}) = 0$.
Notwithstanding, let us remark that the explicit inclusion of a term proportional to the axial gauge condition in the interpolation which we consider would be somewhat awkward and undesirable.
Indeed the parameter weighting the axial gauge term would not have the same dimension as
$\alpha$ and $\beta$. This would in turn
complicate\footnote{The complications are analogous to massive perturbations of Conformal Field Theory
where simple rational functions are deformed into Bessel functions.} the tensor structure and the
explicit functional $x$-dependence of the two-point autocorrelator in $x$-space in such a generalized
interpolating gauge in a nasty way. Fortunately the above described procedure bypasses these
annoyances regarding the two-point autocorrelator of $A$. Yet we shall note in addition that, as a consequence
of the difference in dimensions mentioned above, the two-point correlator of $A$ and $B$ (unexplicited in this
letter) have different expressions in the axial gauge and in the so-called axial limit above.

\section{Observables and geometric interpretations}
\label{Observables}

As discussed in \cite{GPT1}, the relevant quantity to compute the expectation values of observables is:
\begin{equation}
\left(\mathcal{J}_{1},\mathfrak{P}\mathcal{J}_{2}\right)
=\int_{\R_{x}^{4l+3}}d^{4l+3}x\int_{\R_{y}^{4l+3}}d^{4l+3}y\mathcal{J}_{1}^{\mu_{1}\cdots\mu_{2l+1}}\left(x\right)
\mathfrak{P}_{\mu_{1}\cdots\mu_{2l+1}\nu_{1}\cdots\nu_{2l+1}}\left(x-y\right)
\mathcal{J}_{2}^{\nu_{1}\cdots\nu_{2l+1}}\left(y\right)
\end{equation}
$\mathcal{J}_{1}$ and $\mathcal{J}_{2}$ being the de Rham currents \cite{dR} associated to the observables (Wilson loops).

In the covariant gauge this formula is precisely the Gauss linking formula generalized to dimension $4l+3$.
Its classical geometric interpretation in terms of zodiacal $4l+2$-sphere and solid angle has been discussed
in some details in Appendix B of ref. \cite{GPT1}.
This extends to the general interpolating gauge. The linking formula in the interpolating gauge reads explicitely:
\begin{equation}
\left(\mathcal{J}_{1},\mathfrak{P}\mathcal{J}_{2}\right)
= \frac{1}{S_{4l+2}} \oint_{\gamma_1}\oint_{\gamma_2}
\frac{\reallywidetilde{[\widetilde{e}(x,y) , dx , dy]}}{\widetilde{\abs{x-y}}^{4l+2}}  \,\, , \,\,
S_n = \frac{2\pi^{\frac{n+1}{2}}}{\Gamma \left( \frac{n+1}{2} \right)}
\end{equation}
(recall that $\widetilde{\abs{.}}$ is the distance associated with the anisotropic $\widetilde{g}$ metric). The antisymmetric product
\begin{equation}
 \widetilde{[e,a,b]} =
 \frac{\sqrt{\abs{\widetilde{g}}}}{(2l+1)!^2}
 \epsilon_{\sigma \mu_1...\mu_{2l+1}\nu_1...\nu_{2l+1}}
 e^{\sigma}a^{\mu_1}...a^{\mu_{2l+1}}b^{\nu_1}...b^{\nu_{2l+1}}
\end{equation}
is closely related to the Riemannian volume form for this metric and the vector $\widetilde{e}(x,y)$ is defined by $\widetilde{e}(x,y) = \frac{x-y}{\widetilde{\abs{x-y}}}$.
$S_n$ is the area of the $n$-sphere $S^n$. This linking formula has the following twofold interpretation.

In the original Euclidean space with isotropic metric, $\widetilde{e}$ defines the Gauss map
from the product of spheres $S^{2l+1} \times  S^{2l+1}$ onto the zodiacal $(4l+2)$-ellipsoid
${\cal E}_{\alpha \gamma}$, defined by $\widetilde{\abs{x}}=1$. The degree of the Gauss map, {\it e.g.} the number of times the
ellipsoid is covered, is the linking number. Accordingly, the above linking formula is the one for the solid angle related to the ellipsoid.

The linking number, or the degree of the Gauss map, is an integer and as such it cannot vary along a
continuous change of the interpolating parameters $\alpha$ and $\gamma$.
Indeed, with $\alpha \gamma > 0$, the linking formula can be reinterpreted in a Riemannian space with
constant anisotropic metric $\widetilde{g}$ where the Gauss map $\widetilde{e}$ image is the
zodiacal $(4l+2)$-sphere and the linking formula is exactly the Gauss formula.


\vspace{0.3cm}

In the Coulomb gauge ($\gamma=0$) and in the so-called axial limit ($\alpha=0$), we find also
well-known linking formulae, in terms of transverse intersection in the Coulomb gauge, in terms
of counting of oriented crossings in the axial limit.
The linking formula in the Coulomb gauge may also be interpreted in zodiacal terms.
Yet, as a counterpart of the distributional {\it i.e.} singular character of the limit,
the corresponding ``limit zodiacs'' are dimensionaly shrunked: the zodiac degenerates
into the equatorial $(4l+1)$-sphere.
In the axial limit, the zodiac degenerates into a pair of ``antipodal'' poles.

\vspace{0.3cm}

It shall be stressed that linking invariants
are characteristic, not of loops {\em per se} but of equivalence classes of loops w.r.t. isotopy.
Notwithstanding, the computation of linking invariants using {\em representatives} of classes demands
restrictions on these representatives in order for the calculation to be well-defined. This is most simply
visualised in the three dimensional case in what follows.
In covariant gauge there is no more restriction than the general prerequisite that the loops shall not
(self-) intersect.

\begin{figure}[!ht]
\centering
\includegraphics[width =4cm]{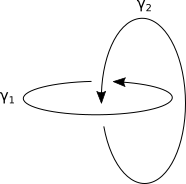}
\caption{Typical configuration in covariant gauge.}
\end {figure}

The situations for the Coulomb gauge and the axial limit are more demanding due to the
nature of the autocorrelators (\ref{coulombprop}) and (\ref{axialprop}), as products of delta distributions with non transverse
supports lead to pathologies.

Let us assume that one of the loops lies in a plane to keep things simple.
In Coulomb gauge, a pathology occurs if a continuous arc of the other loop happens to be contained in this plane:
\begin{figure}[!ht]
\centering
\begin{minipage}[t]{7cm}
\centering
\includegraphics[width =6cm]{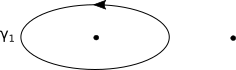}
\subcaption{Typical configuration.}
\end{minipage}
\hspace{2cm}
\begin{minipage}[t]{7cm}
\centering
\includegraphics[width=6cm]{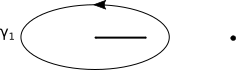}
\subcaption {Pathological configuration.}
\end {minipage}
\caption{Configurations in Coulomb gauge.}
\end {figure}
so that tranverse intersection is not well defined. Accordingly, for such a configuration,
the Gauss map becomes an application from a two-dimensional space\footnote{This space is a cylinder in the example of the configuration depicted in figure 2b.} to a one-dimensional space.
Hence the notion of degree for such an application is meaningless.

In the axial limit, a pathology occurs if
a continuous arc of projection of the other loop onto the plane containing the first loop
happens to coincide with an arc of the first loop:
\begin{figure}[!ht]
\centering
\begin{minipage}[t]{7cm}
\centering
\includegraphics[width=6cm]{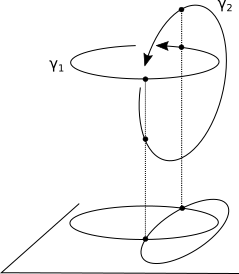}
\subcaption{Typical configuration.}
\end{minipage}
\hspace{1cm}
\begin{minipage}[t]{7cm}
\centering
\includegraphics[width=6cm]{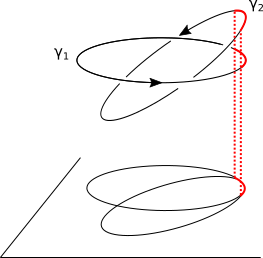}
\subcaption{Pathological configuration.}
\end {minipage}
\caption{Configurations in axial limit gauge.}
\end {figure}
and crossing is not properly defined. Accordingly, the Gauss map is an application from a
one-dimensional space to a zero-dimensional space and its degree is not defined.

\section{Conclusion}

The covariant and Coulomb gauges as well as the axial limit can be regarded as limits of the interpolating gauge. In the regime $\alpha \gamma > 0$ the interpolating gauge in the Euclidean metric is equivalent to the covariant gauge in an anisotropic constant metric. In this metric the Gauss sphere becomes an ellipsoid. This leads to geometric interpretations for the previous limits even in the singular cases.

Our study based on a two-parameter interpolating gauge extends to a generic interpolating gauge equivalent to the covariant gauge in a completely anisotropic constant Riemannian metric $\widetilde{g} = diag(\alpha_1 , \cdots , \alpha_{4l+3})$. A wider collection of gauge fixings is obtained as singular limits of the corresponding interpolating gauge for subsets of vanishing $\alpha_i$'s. This provides a corresponding collection of propagators hence of expressions of the linking number of closed $2l+1$ dimensional submanifolds in $\mathbb{R}^{4l+3}$. Each of these limits has its own set of pathological configurations.

In three dimensions we can make \textit{formal} contact between the present work and \cite{LLSS} as follows. The gauge fixing of \cite{LLSS} implies the gauge condition $\mathcal{F}\left(A\right) =  \left(\partial^{\mu} A_{\nu}\right) + \zeta \left(n^{\nu}\partial_{\nu}\right)\left(n^{\mu}A_{\mu}\right) = \partial_{\mu} [ \left(\eta^{\mu\nu} +\zeta n^\mu n^\nu\right) A_{\nu} ]$ where $\eta^{\mu\nu}$ is the Minkowski metric, $n^\mu$ is a lightlike vector and $\zeta$ a real parameter.
In our approach this would correspond to considering a (inverse) metric $\widetilde{g}^{\mu\nu} = \eta^{\mu\nu} + \zeta n^\mu n^\nu$ which turns out to be also of Minkowski type for all $\zeta$. In contrast in this article we restricted ourselves to (rigid) \textit{Riemannian} metrics by means of the condition $\alpha \gamma > 0$ (and its generalisation $\alpha_1 , \cdots , \alpha_{4l+3}$ all positive). Whereas Riemannian metrics provide distances which make contact between geometry and topology, pseudo-Riemannian metrics do not. The relation between CS in pseudo-Riemannian background, appearing through a gauge fixing procedure, and topological features is obscure. We do not intend to elaborate any further on this issue in the present article.

The results presented here have a significance larger than the abelian Chern-Simons case. In fact, the propagator is not specific to this theory. As it is formally the inverse of the operator of the quadratic part of the action, it corresponds also to the propagator of non-abelian Chern-Simons case. The operator of the quadratic part comes out also in abelian and non-abelian BF theories which thus share the same propagator as abelian and non-abelian Chern-Simons theories.

\section{Appendix}

\subsection{Solution of the inversion equation}

We use in what follows the following identity:
\begin{equation}
\left(-1\right)^{r}\left(\widetilde{\hodge }\widetilde{K}\wedge\widetilde{\hodge }\widetilde{N}\wedge-\widetilde{N}\wedge\widetilde{\hodge }\widetilde{K}\wedge\widetilde{\hodge }\right)\widetilde{\omega}^{\left(r\right)}
= \left(n\widetilde{\cdot} k\right) \widetilde{\omega}^{\left(r\right)}
\end{equation}
in the case of the metric $\widetilde{g}$.

We solve now the following system:
\begin{align}
-i\left(\widetilde{K}\wedge\hat{\widetilde{P}}^{\left(2l+1,2l+2-j\right)}_{1,j}+\widetilde{K}^{\left(1\right)}\wedge\hat{\widetilde{P}}^{\left(2l,2l+2-j\right)}_{2,j}\right)
&=\delta_{1,j}\id \mbox{ for } j\in\intervalleentier{1}{2l+2}\\
i\left(\left(-1\right)^{i}\widetilde{\hodge }\widetilde{K}\wedge\widetilde{\hodge }\hat{\widetilde{P}}^{\left(2l+3-i,2l+2-j\right)}_{i-1,j}-\widetilde{K}\wedge\hat{\widetilde{P}}^{\left(2l+1-i,2l+2-j\right)}_{i+1,j}\right)
&=\delta_{i,j}\id \mbox{ for } j\in\intervalleentier{1}{2l+2}, i\in\intervalleentier{2}{2l+1}\\
i\left(\widetilde{\hodge }K\widetilde{\hodge }\wedge\hat{\widetilde{P}}^{\left(1,2l+2-j\right)}_{2l+1,j}\right)
&=\delta_{2l+2,j}\id \mbox{ for } j\in\intervalleentier{1}{2l+2}
\end{align}

First, we notice that:
\begin{equation}
-i\widetilde{\hodge }\widetilde{K}\wedge\widetilde{\hodge }\hat{\widetilde{P}}^{\left(1,2l+1\right)}_{2l+1,1}=0,
\end{equation}
but:
\begin{equation}
\widetilde{\hodge }\widetilde{K}\wedge\hat{\widetilde{P}}^{\left(3,2l+1\right)}_{2l-1,1}+\widetilde{K}\wedge\hat{\widetilde{P}}^{\left(1,2l+1\right)}_{2l+1,1}=0.
\end{equation}
So, operating with $\widetilde{\hodge }\widetilde{K}\wedge\widetilde{\hodge }$ on this last equation, we get:
\begin{equation}
\widetilde{\hodge }\widetilde{K}\wedge\widetilde{K}\wedge\hat{\widetilde{P}}^{\left(3,2l+1\right)}_{2l-1,1}-\widetilde{\hodge }\widetilde{K}\wedge\widetilde{\hodge }\widetilde{K}\wedge\hat{\widetilde{P}}^{\left(1,2l+1\right)}_{2l+1,1}
=-\widetilde{\hodge }\widetilde{K}\wedge\widetilde{\hodge }\widetilde{K}\wedge\hat{\widetilde{P}}^{\left(1,2l+1\right)}_{2l+1,1}
=0,
\end{equation}
so:
\begin{equation}
-\left(-k\widetilde{\cdot}k-\widetilde{K}\wedge\widetilde{\hodge }\widetilde{K}\wedge\widetilde{\hodge }\right)\hat{\widetilde{P}}^{\left(1,2l+1\right)}_{2l+1,1}
=k\widetilde{\cdot}k \hat{\widetilde{P}}^{\left(1,2l+1\right)}_{2l+1,1}=0,
\end{equation}
and thus:
\begin{equation}
\hat{\widetilde{P}}^{\left(1,2l+1\right)}_{2l+1,1} = 0.
\end{equation}
This procedure can be repeated so that $\hat{\widetilde{P}}^{\left(2l+1-2i,2l+1\right)}_{2i+1,1} = 0$ until $i=1$ and we get at the next step that:
\begin{equation}
\widetilde{\hodge }\widetilde{K}\wedge\widetilde{\hodge }\hat{\widetilde{P}}^{\left(2l+1,2l+1\right)}_{1,1}=0.
\end{equation}
We now consider the equation:
\begin{equation}
-i\left(\widetilde{\hodge }\widetilde{K}\wedge\hat{\widetilde{P}}^{\left(2l+1,2l+1\right)}_{1,1}-\widetilde{K}\wedge\hat{\widetilde{P}}^{\left(2l,2l+1\right)}_{2,1}\right)=\id,
\end{equation}
on which we operate with $\widetilde{\hodge }\widetilde{K}\wedge$ so that we obtain:
\begin{equation}
-i\widetilde{\hodge }\widetilde{K}\wedge\widetilde{\hodge }\widetilde{K}\wedge\hat{\widetilde{P}}^{\left(2l+1,2l+1\right)}_{1,1}
=\widetilde{\hodge }\widetilde{K}\wedge.
\end{equation}
Thus:
\begin{equation}
-i\left(k\widetilde{\cdot}k-\widetilde{K}\wedge\widetilde{\hodge }\widetilde{K}\wedge\widetilde{\hodge }\right)\hat{\widetilde{P}}^{\left(2l+1,2l+1\right)}_{1,1}
=-i k\widetilde{\cdot}k\hat{\widetilde{P}}^{\left(2l+1,2l+1\right)}_{1,1}
=\widetilde{\hodge }\widetilde{K}
\end{equation}
and finally:
\begin{equation}
\hat{\widetilde{P}}^{\left(2l+1,2l+1\right)}_{1,1}
=-i\widetilde{\hodge }\frac{\widetilde{K}\wedge\cdot}{k\widetilde{\cdot}k}
\end{equation}

\subsection{Computation of the inverse Fourier transformation}

\noindent
The inverse Fourier transform
$\widetilde{P}^{\left(2l+1,2l+1\right)}_{1,1}$ of
$\hat{\widetilde{P}}^{\left(2l+1,2l+1\right)}_{1,1}$ is defined by:
\begin{align}
\widetilde{P}_{\mu_{1} \cdots \mu_{4l+1} \nu_{1} \cdots \nu_{4l+1}}
\left( x \right)
&=\frac{\abs{\widetilde{g}}}{\left(2\pi\right)^{4l+3}}
\int\int_{\R \times \R^{4l+2}}
\hat{\widetilde{P}}_{\mu_{1} \cdots \mu_{4l+1} \nu_{1} \cdots\nu_{4l+1}}
\left( k \right)
e^{i k\tilde{\cdot} x}d k_{\parallel} d^{4l+2} k_{\perp}
\label{e72-1}
\end{align}
It takes the explicit form:
\begin{align}
\widetilde{P}_{\mu_{1} \cdots \mu_{4l+1} \nu_{1} \cdots \nu_{4l+1}}
\left( x \right)
&=-\frac{i\sqrt{\abs{\widetilde{g}}}}{\left(2l+1\right)!}
\varepsilon_{\mu_{1}\cdots\mu_{4l+1}\nu_{1}\cdots\nu_{4l+1}\rho}
\frac{\abs{\widetilde{g}}}{\left(2\pi\right)^{4l+3}}
\int_{\R} d k_{\parallel} \int_{\R^{4l+2}} d^{4l+2} k_{\perp}
\frac{k^{\rho}}
{\left(
  \frac{1}{\alpha} k_{\perp}^{2}+\frac{1}{\gamma} k_{\parallel}^{2}
 \right)}
e^{i k\tilde{\cdot} x}
\label{e72-2}
\end{align}
Using Schwinger's ``proper time'' parametrisation
\begin{equation}
\frac{1}{\frac{1}{\alpha} k_{\perp}^{2}+ \frac{1}{\gamma} k_{\parallel}^{2}}
=
\int_{0}^{+\infty} d\tau \,
e^{- \tau \left( \frac{1}{\alpha} k_{\perp}^{2} +
     \frac{1}{\gamma} k_{\parallel}^{2} \right)}
\nonumber
\end{equation}
the integral in the r.h.s. of eq. (\ref{e72-2}) reads:
\begin{equation}
\int_{\R} d k_{\parallel} \int_{\R^{4l+2}} d^{4l+2} k_{\perp}
\frac{k^{\rho}}
{\left(
  \frac{1}{\alpha} k_{\perp}^{2}+\frac{1}{\gamma} k_{\parallel}^{2}
 \right)} \,
e^{i \, k \tilde{\cdot} x}
=\int_{0}^{+\infty} d\tau
\int_{\R} d k_{\parallel} \int_{\R^{4l+2}} d^{4l+2} k_{\perp}
k^{\rho}
e^{i \left( \frac{1}{\alpha}k_{\perp}x_{\perp} +
           \frac{1}{\gamma}k_{\parallel}x_{\parallel}
     \right)}
e^{-\tau \left( \frac{1}{\alpha} k_{\perp}^{2} +
                \frac{1}{\gamma} k_{\parallel}^{2}
         \right)}
\label{e72-3}
\end{equation}
We set:
\begin{center}
\begin{tabular}{lcr}
$q_{\parallel}= \frac{1}{\sqrt{\gamma}}k_{\parallel}$ &
$\;\; \mbox{and} \;\;$ & $q_{\perp}=\frac{1}{\sqrt{\alpha}}k_{\perp},$ \\
$z_{\parallel}= \frac{1}{\sqrt{\gamma}}x_{\parallel}$ &
$\;\; \mbox{and} \;\;$ & $z_{\perp}=\frac{1}{\sqrt{\alpha}}x_{\perp},$ \\
\end{tabular}
\end{center}
and rewrite the r.h.s. of eq. (\ref{e72-3}) as:
\begin{align}
&\int_{\R} d k_{\parallel}
\int_{\R^{4l+2}} d^{4l+2} k_{\perp}
\frac{k^{\rho}}
{\left(
  \frac{1}{\alpha} k_{\perp}^{2}+\frac{1}{\gamma} k_{\parallel}^{2}
 \right)} \,
e^{i \, k \tilde{\cdot} x}
\notag\\
& \;\;\;\;\;\;\;\;\;\;\;\;\;\;\;\;\;\;\;\;\;\;\;\;\;\;\;\;\;
 = \int_{\R} \sqrt{\gamma} \, d q_{\parallel}
\int_{\R^{4l+2}} \sqrt{\alpha}^{4l+2} \, d^{4l+2} q_{\perp}
 \int_{0}^{+\infty} d \tau \left(\sqrt{\alpha}q_{\perp}^{\rho}+\sqrt{\gamma}q_{\parallel}^{\rho}\right)
e^{i \left(q_{\perp} z_{\perp} + q_{\parallel} z_{\parallel} \right) -
\tau \left(q_{\perp}^2 + q_{\parallel}^2 \right)}
\label{e72-4}
\end{align}
We perform the gaussian integration over $q$ first, conveniently
casting the result in the form:
\begin{align}
\int_{\R} \frac{d k_{\parallel}}{2 \pi}
\int_{\R^{4l+2}} \frac{d^{4l+2} k_{\perp}}{\left( 2 \pi\right)^{4l+2}}
\frac{k^{\rho}}
{\left(
  \frac{1}{\alpha} k_{\perp}^{2}+\frac{1}{\gamma} k_{\parallel}^{2}
 \right)} \,
e^{i \, \widetilde{k \cdot x}}
&= - i \, \left( 4 \pi \right)^{-\frac{4l+3}{2}} \,
\sqrt{\gamma \, \alpha^{4l+2}}
\notag\\
&\qquad \times
\left\{
 \sqrt{\alpha} \, \frac{\partial}{\partial z_{\perp}^{\rho}}
 +
 \sqrt{\gamma} \, \frac{\partial}{\partial z_{\parallel}^{\rho}}
\right\}
\int_{0}^{+\infty} d\tau \, \tau^{-\frac{1}{2}(4l+3)} e^{-\frac{z^{2}}{4\tau}}
\label{e72-5}
\end{align}
The last integral over $\tau$ is readily performed setting $\tau = 1/u$:
\begin{align}
\int_{0}^{+\infty} d\tau \, \tau^{-\frac{1}{2}(4l+3)} e^{-\frac{z^{2}}{4\tau}}
&=
\Gamma \left( \frac{4l+1}{2} \right) \, 2^{4l+1} \,
\left( z^{2} \right)^{-\frac{4l+1}{2}}
\end{align}
so that:
\begin{align}
\int_{\R} \frac{d k_{\parallel}}{2 \pi}
\int_{\R^{4l+2}} \frac{d^{4l+2} k_{\perp}}{\left( 2 \pi\right)^{4l+2}}
\frac{k^{\rho}}
{\left(
  \frac{1}{\alpha} k_{\perp}^{2}+\frac{1}{\gamma} k_{\parallel}^{2}
 \right)} \,
e^{i \, k \tilde{\cdot} x}
&=
i\frac{\Gamma\left(\frac{4l+3}{2}\right)}{2 \, \pi^{\frac{4l+3}{2}}} \,
\sqrt{\gamma \, \alpha^{4l+2}} \,
\frac{\sqrt{\alpha}z_{\perp}^{\rho}+\sqrt{\gamma}z_{\parallel}^{\rho}}
     {\left(z^{2}\right)^{\frac{4l+3}{2}}}
\end{align}
Finally, with $|\widetilde{g}| = 1/(\gamma \, \alpha^{4l+2})$ and
$\sqrt{\alpha}z_{\perp}^{\rho}+\sqrt{\gamma}z_{\parallel}^{\rho}=x^{\rho}$,
we get:
\begin{equation}
\widetilde{P}_{\mu_{1}\cdots\mu_{4l+1}\nu_{1} \cdots \nu_{4l+1}} \left(x\right)
=\frac{1}{\left(2l+1\right)!} \,
\frac{\Gamma \left( \frac{4l+3}{2} \right)}{2 \, \pi ^{\frac{4l+3}{2}}}
\frac{1}{\gamma \, \alpha^{4l+2}}
\varepsilon_{\mu_{1}\cdots\mu_{4l+1}\nu_{1}\cdots\nu_{4l+1}\rho}
\frac{x^{\rho}}
{\left(\frac{1}{\gamma}x_{\parallel}^{2}+\frac{1}{\alpha}x_{\perp}^{2}\right)^{\frac{4l+3}{2}}}
\end{equation}


\begin{thebibliography}{MT02}

\bibitem{W}
E. Witten, \textit{Quantum field theory and the Jones polynomial}, Comm. Math. Phys. \textbf{121}, Number 3, 351 (1989)

\bibitem{GT1}
E. Guadagnini and F. Thuillier, {\it Deligne-Beilinson Cohomology and Abelian Link Invariants}, SIGMA \textbf{4}, 078 (2008).

\bibitem{GPT1}
L. Gallot, É. Pilon, F. Thuillier, \textit{Higher dimensional abelian Chern-Simons theories and their link invariants}, JMP \textbf{54}, 022305 (2013)

\bibitem{BZ}
L. Baulieu, D. Zwanziger, \textit{Renormalizable Non-Covariant Gauges and Coulomb Gauge Limit}, Nucl. Phys. \textbf{B548}, 527 (1999)

\bibitem{DGLSSSTV}
D. Dudal, J.A. Gracey, V.E.R. Lemes, R.F. Sobreiro, S.P. Sorella, R. Thibes, H. Verschelde, \textit{Remarks on a class of renormalizable interpolating gauges}, JHEP \textbf{0507}, 059 (2005)

\bibitem{LLSS}
K. Landsteiner, M. Langer, M. Schweda, S. P. Sorella, \textit{Interpolating gauge fixing for Chern-Simons theory}, Phys. Lett. \textbf{B337}, 294 (1994)

\bibitem{dR}
G. de Rham, {\it Variétés Differentiables, Formes, Courants, Formes Harmoniques}, Hermann (Paris, 1955).


\end{thebibliography}
\end{document}